\documentstyle[12pt,russian]{article}
\begin{document}

\begin{center}
{\large\bf GRAVITON MASS, 
QUINTESSENCE AND OSCILLATORY CHARACTER OF THE UNIVERSE EVOLUTION}

\vspace*{0.2cm}
{\bf S.S. Gershtein$^{1}$, A.A. Logunov$^{2}$, M.A. Mestvirishvili,\\
N.P. Tkachenko$^{3}$}\\
{\small \it Institute for High Energy Physics,\\
Protvino, Russia}
\end{center}

\vspace*{2cm}
\noindent
{\small It is shown that using the relativistic field theory of gravity
(RTG) and measured value of  $\Omega_{\mbox{{\small tot}}}$
one can obtain the upper limit on the graviton mass with  95\%C.L.:
$m\leq 1.6\cdot 10^{-66}$ [g]; within the $(1\sigma)$ range 
its probable value is $m_{g}= 1.3\cdot 10^{-66}$ [g]. It is pointed out
that according to RTG the presence of the quintessence is necessary
to explain the Universe accelerated expansion. Experimental data
on the Universe age and dark matter density allow one to determine
the range of possible values of the $\nu$ parameter in the  
equation of quintessence state and indicate characteristic time, which corresponds
to the beginning and cessation  of the accelerated expansion epoch, as well as
the time period of the maximal expansion, which corresponds to the 
half-period  of the oscillatory evolution of the Universe.}

\vspace*{4cm}
\begin{verbatim}
gershtein@mx.ihep.su
logunov@mx.ihep.su
tkachenkon@mx.ihep.su
\end{verbatim}

\newpage

\setcounter{page}{1}

\begin{center}
{\large\bf GRAVITON MASS, 
QUINESSENCE AND OSCILLATORY CHARACTER OF THE UNIVERSE EVOLUTION}

\vspace*{0.2cm}
{\bf S.S. Gershtein\footnote{e-mail: gershtein@mx.ihep.su}, A.A.
Logunov\footnote{e-mail: logunov@mx.ihep.su}, M.A. Mestvirishvili,\\
N.P. Tkachenko\footnote{e-mail: tkachenkon@mx.ihep.su}}\\
{\small \it Institute for High Energy Physics,\\
Protvino, Russia}
\end{center}

\underline{Abstract}

\noindent
{\small It is shown that using the relativistic field theory of gravity
(RTG) and measured value of  $\Omega_{\mbox{{\small tot}}}$
one can obtain the upper limit on the graviton mass with  95\%C.L.:
$m\leq 1.6\cdot 10^{-66}$ [g]; within the $(1\sigma)$ range 
its probable value is $m_{g}= 1.3\cdot 10^{-66}$ [g]. It is pointed out
that according to RTG the presence of the quintessence is necessary
to explain the Universe accelerated expansion. Experimental data
on the Universe age and dark matter density allow one to determine
the range of possible values of the $\nu$ parameter in the  
equation of quintessence state and indicate characteristic time, which corresponds
to the beginning and cessation
 of the accelerated expansion epoch, as well as
the time period of the maximal expansion, which corresponds to the 
half-period  of the oscillatory evolution of the Universe.}

\section*{\S 1. Introduction}

The discovery of the Universe accelerated expansion \cite{1}~$\div$ \cite{3}
has forced to revise many established ideas on the Universe content and 
character of its evolution. One of the popular explanations
of the accelerated expansion is the assumption on the presence 
of the cosmological constant, $\Lambda$, that is equivalent to
the existence of nonvanishing vacuum energy, 
$\varepsilon_0$, and related negative pressure
$P_0 =-\varepsilon_0$ (\cite{4}~$\div$ \cite{8}). Such an assumption
leads to unlimited inflationary expansion of the Universe
(which rate, however, is at least by  60 orders of magnitude less than 
the initial inflationary expansion from the Plank scales assumed to solve
the horizon problem and explain the flat geometry of the three-dimensional
space). Another alternative explanation of the observed accelerated 
expansion involves the hypothesis of the existence of the special
substance in the Universe  -- the quintessence \cite{9}~$\div$ \cite{11}
with equation of state as follows
\begin{equation}
\boldmath{
P_q=-(1-\nu)\varepsilon_q ~~~
\left(
0<\nu < \frac{2}{3}
\right)
\;,
}
\label{eq:1}
\end{equation}
where $\varepsilon_q$ and $P_q$ are the density of energy and
quintessence pressure, correspondingly. 
 
Relativistic field theory of gravity ('ƒ) \cite{12,13}, which consider
the gravitational field as a physical field in the Minkowsky space,
is inconsistent with unlimited expansion of the Universe.
Thus, as it was shown by V.P. Kalashnikov \cite{14}, the existence 
of quintessence (\ref{eq:1}) is essential to explain the observed
accelerated expansion of the Universe in framework of RTG. In this case, 
according to RTG,
the cyclic evolution of the Universe will take place. The importance
of this question and appearance of new experimental data impel us to 
come back to this problem. 

In this paper it is shown that under the assumption 
of the quintessence existence the relativistic theory of gravity (RTG), 
in which there is no cosmological singularity and the 
flat character of the three-dimension space geometry is 
predicted unambiguously, leads to the fact that 
at  $\nu >0$ the current acceleration of the Universe expansion further
should be changed into decreasing and then the expansion will stop;
after that the ``compression'' up to some minimal value of the scale factor will start, then again the expansion cycle will take place.
In \S 2 the basics of RTG are summarized. In
\S3  the  consequences for evolution of the uniform and isotropic 
Universe following from these basic laws are scoped. In \S 4 we use recent experimental estimate on the cosmological parameter  $\Omega_{\mbox{{\small tot}}}$ to put the  95\% C.L. upper limit on the graviton mass, and its probable value is determined
within the error range of  $(1\sigma)$. In \S 5 using this value
we confine the region of the allowed values of the  $\nu$ parameter, which
agrees with the date on current age of the Universe and other cosmological
parameters. In \S 6 we estimate the characteristic time, corresponding to
the beginning and the cessation
 of currently observed expansion acceleration, as well as the possible period of the Universe oscillation.

\section*{\S 2. Basics of RTG}

RTG starts from the assumption that the gravitational field, as all other fields, develops in the Minkowsky space and that the tensor of the energy-momentum of all matter fields, including the gravitational field,
is the source of this field. Such approach is concordant with
modern gauge theories of the electroweak interactions and QCD, where
conserving charges and their currents are the source of the vector fields. As the energy-momentum tensor is chosen to be the source of the gravitational field, the gravitational field itself should be described by
symmetrical tensor of the second rank, $\varphi^{\mu\nu}$. Further this gives rise to a ``geometrization'' of the theory. The initial set of the RTG equations has the form \cite{12,13} $(\hbar = c = 1)$:
\begin{equation}
(\gamma^{\alpha\beta} D_\alpha D_\beta + m^2_g) \cdot
\tilde\varphi^{\mu\nu} = 16\pi G  t^{\mu\nu},
\label{eq:2}
\end{equation}
\begin{equation}
D_\nu \tilde\varphi^{\mu\nu}=0,
\label{eq:3}
\end{equation}
where $D_\alpha$ is the covariant derivative in the Minkowsky space with
metric tensor $\gamma_{\alpha\beta}$, 
$\tilde \varphi^{\mu\nu}$ and $t^{\mu\nu}$ are the densities of the gravitational field and  total energy-momentum tensor, correspondingly:
$$
\tilde\varphi^{\mu\nu} =\sqrt{-\gamma}\cdot\varphi^{\mu\nu};~~
\gamma = \det (\gamma_{\mu\nu})= \det (\tilde \gamma^{\mu\nu}),~~
t^{\mu\nu} = -2 \cdot \frac{\delta L}{\delta\gamma_{\mu\nu}} ,
$$
where $L$ is the density of the matter and gravitational field Lagrangian. 
Eq. (\ref{eq:3}) guarantees the conservation of the total energy-momentum tensor, singles out polarization states corresponding to the gravitons with the spin  2 and 0, and excludes the states with the spin 1 ¨
0$^\prime$ (analogously to the Lorentz condition, which excludes the photon with the spin 0). For equation set  
(\ref{eq:2})-(\ref{eq:3}) to follow from the minimal action principle, i.e. 
for it to result from the Euler equations\footnote{
Here $L_M(\tilde\gamma^{\mu\nu}, \tilde\varphi^{\mu\nu},
\tilde\varphi_k)$ is the matter Lagrangian density, which corresponds to
the motion of the matter field $\varphi_k$ in the gravitational field,  $L$ 
is the density of the full Lagrangian, which includes the Lagrangian
of the gravitational field $L_g$ itself.}:
\begin{equation}
\frac{\delta L}{\delta \tilde\varphi^{\mu\nu}}=0,\;\;
\frac{\delta L_M}{\delta \tilde\varphi_k} = 0 \;,
\label{eq:5}
\end{equation}
it is necessary and sufficient that the density of the $\tilde\varphi^{\mu\nu}$ tensor and density of the Minkowsky space metric tensor $\tilde\gamma^{\mu\nu}$ should enter into the matter Lagrangian in combination \cite{12,13}:
$$
\tilde\varphi^{\mu\nu} +\tilde\gamma^{\mu\nu}=\tilde g^{\mu\nu};\;\;
\tilde g^{\mu\nu}=\sqrt{-g}\cdot g^{\mu\nu};\;\;
g=\det (\tilde g^{\mu\nu})=\det g_{\mu\nu}\;.
$$
Thus
$$
L=L_g +L_M (\tilde g^{\mu\nu}, \tilde\varphi_k\;,
$$
and the matter motion in the gravitational field looks like, as it
would appear in the effective Riemann space with the metrics $g_{\mu\nu}$. It should be noted that all the crucial changes as compared with 
the Einstein general theory of relativity appear in the RTG
due to the fact that the gravitational field is considered as the
physical field in the Minkowsky space. Namely this approach
essentially leads to the presence of the graviton. Gravity equations
with nonvanishing graviton mass have been exploited early as well (see, for instance, \cite{15}). However, they were written for inertial frames of reference only, since the special theory of relativity has been considered to be valid only for such frames. Therefore these equations naturally 
turned out not to be generally covariant, and due to this fact were not
treated seriously.
In its turn RTG takes into account the fact that in the Minkowsky space 
one can use any frames of reference, including accelerated ones, in which metric coefficients $\gamma_{\mu\nu}$ form the tensor with respect to 
arbitrary coordinate transformation. That is why Eqs.
(\ref{eq:2}) and (\ref{eq:3}) are generally covariant. 

The necessity to introduce the nonvanishing graviton mass
in the gravity field theory is caused by the fact in the case of its absence the gravitational field  $\varphi^{\mu\nu}$ (with the
total energy-momentum tensor to be its source) obeys the group of gauge transformation \cite{12, 13} (see also
\cite{16}), which presence leads to  the fact that some 
physical observables (metric tensor of the effective Riemann space
and its curvature among them) are dependent on the choice of the gauge. 
Introduction of the graviton mass breaks the gauge group and by this
guarantees the independence of the physical observables on the   
any arbitrariness, meanwhile preserving the general covariant property 
of the gravity equations.
The structure of the term in the gravitational field Lagrangian, which
breaks gauge arbitrariness of the gravitational field by introducing
a nonvanishing graviton mass, was unambiguously derived in papers
 \cite{12, 13}. As the result, the equations of the gravitational field and matter takes the form
\begin{equation}
R^{\mu\nu} - \frac{1}{2} g^{\mu\nu}R
+ \frac{1}{2}
\left (
\frac{m_gc}{\hbar}
\right )^2
\left [
g^{\mu\nu} +
\left (
g^{\mu\alpha}g^{\nu\beta}
-\frac{1}{2}
g^{\mu\nu}g^{\alpha\beta}
\right )
\gamma_{\alpha\beta}
\right ]
= 8\pi GT^{\mu\nu}\; , \label{eq:8}
\end{equation}
\begin{equation}
D_\mu \tilde g^{\mu\nu} = 0, \label{eq:9}
\end{equation}
where $R^{\mu\nu}$ and $R$ are the corresponding curvatures in the 
effective Riemann space, and $T^{\mu\nu}$ is the energy-momentum tensor
of the matter in the effective Riemann space
$$
\sqrt{-g} \cdot T^{\mu\nu} =
-2\cdot \frac{\delta L_M}{\delta g_{\mu\nu}};
$$
Eqs. (\ref{eq:8})~-- (\ref{eq:9}) are generally covariant with respect to arbitrary coordinate transformations and form-invariant with respect tj the Lorentz transformations. Due to  $m_g\ne 0$, the connection of the effective Riemann space with the metrics of the original Minkowsky space
$\gamma_{\alpha\beta}$ in Eq. (\ref{eq:8})is retained. 

{\bf Eqs. (\ref{eq:8})-(\ref{eq:9}) form the complete set of equations}. It is necessary to stress out that here relation (\ref{eq:9})
is namely the {\bf equation}, which is the consequence of the law of the  total energy-momentum tensor conservation (or, that is equivalent,
from Eq. (\ref{eq:5}) for the matter field), rather than any other additional condition. With current estimates on the probable graviton mass (see \S 4) Eqs. (\ref{eq:8}) and (\ref{eq:9}) are fully agree ¯®«­®áâìî with all relativistic gravitational effects observed in the Solar system. 

\section*{\S 3. Evolution of the uniform and isotropic Universe according to  RTG}

For the uniform and isotropic Universe the interval between events 
in the effective Riemann space can be represented in the  {\em Freedman-Robertson-Walker} metrics:
\begin{equation}
ds^2=U(t)\cdot (dx^0)^2-V(t) \cdot
\left [
\frac{dr^2}{1-kr^2}+r^2
(d\Theta^2+\sin^2\Theta d\varphi^2)\right ],
\label{eq:11}
\end{equation}
where $k=1,-1,0$ for the closed (elliptic), open (hyperbolic), and flat
(parabolic) Universe, correspondingly.

Eqs. (\ref{eq:9}) for metrics (\ref{eq:11}) takes the form:
\begin{equation}
\frac{\partial}{\partial t}
\left (
\frac{V^3}{U}\right )
=0,
\label{eq:12}
\end{equation}
\begin{equation}
\frac{\partial}{\partial r}
\left (
r^2\sqrt{1-kr^2}
\right ) -2 r (1-kr^2)^{-1/2}=0 \;.
\label{eq:13}
\end{equation}
It follows from Eq.(\ref{eq:12}) that ${V^3}/{U}=$ const, or
\begin{equation}
V=\beta U^{1/3};\;\;
\beta = \mbox{const}.
\label{eq:14}
\end{equation}
Eq. (\ref{eq:13}) is valid only for $k=0$. 

Thus, from (\ref{eq:9}) it follows immediately that 
{\bf space} {\bf geometry of the Universe has to be flat}
(at that the initial inflationary expansion is not required).
For the first time this result has been noticed in \cite{17}.
The fact that RTG results in the only  (flat) solution ($k= 0$) for 
the uniform and isotropic Universe instead of three possible Freedman solutions is quite natural, as the set of equations
(\ref{eq:8})-(\ref{eq:9}) together with the equation of the state  for
$T^{\mu\nu}$ form a complete set of equations, which has the only solution. 

Introducing the eigen time
$$
d\tau=U^{1/2}\cdot dt
$$
and notation of
$$
a^2(\tau)=U^{1/3},
$$
one can rewrite interval (\ref{eq:11}) in the following form 
\begin{equation}
ds^2=c^2d\tau^2-\beta a^2(\tau) \cdot
\left[
dr^2+r^2
(d\Theta^2+\sin^2\Theta d\Phi^2)
\right] .
\label{eq:17}
\end{equation}
When expressions of (\ref{eq:17}) are used the equations of gravity (\ref{eq:8})
for uniform and isotropic Universe take the form \cite{12,13}
\begin{equation}
\left (
\frac{1}{a}
\frac{da}{d\tau}
\right )^2
=\frac{8\pi G}{3}
\rho -\frac{1}{6}
\left (
\frac{m_gc^2}{\hbar}\right )^2
\left (
1-\frac{3}{2\beta a^2}
+\frac{1}{2a^6}
\right )\; ,
\label{eq:18}
\end{equation}
\begin{equation}
\frac{1}{a}
\frac{d^2a}{d\tau^2}
=-\frac{4\pi G}{3}
\left (
\rho+\frac{3P}{c^2}
\right )
-\frac{1}{6}
\left (
\frac{m_gc^2}{\hbar}\right )^2
\left (
1-\frac{1}{a^6}
\right ) \; , \label{eq:19}
\end{equation}
where $\rho$ and $P$ áare the total density
of all types of matter and the pressure caused by the matter.

The $\beta$ constant, which is determined by Eq. (\ref{eq:14})and enters in Eq. (\ref{eq:18}), has a simple physical meaning. Considering the gravitational field $\varphi^{\mu\nu}$ as a physical field in the Minlowsky space it is necessary to require the fulfillment of the causality principle, which lies in the fact that the trajectory of the particle subjected to the gravitational field action should not leave the limits of the light cone in the Minkowsky space. For interval (\ref{eq:17})
this condition leads to inequality
$$
a^2(\tau) \cdot
\left[
a^4(\tau)-\beta
\right]
\leq 0. 
$$
Thus, the $\beta$ constant determines the maximal value
of the scale multiplier \cite{12,13}.
$$
a^4_{\;\mbox{{\small max}}} =\beta. 
$$
It means that  {\bf according to RTG, the unlimited increase of the scale factor  $a(\tau)$ is not possible}, i.e. the unlimited expansion of the Universe\footnote{We adhere to the traditional notation of the Universe ``expansion'', though, in reality, the Universe is infinite: the $r$ coordinate in interval (17) varies in the range $0<r<\infty$. The increase of the distance between galaxies, detected by means of the red shift and interpreted as the Doppler effect, is the consequence of the fact that
the light signal emission from remote galaxies takes place in the gravitational field, which is more strong, than that one in the  moment
when the observer receives the signal.} is not possible.

According to Eq. (\ref{eq:18}) the nonvanishing graviton mass guarantee the fulfillment of this requirement in the case when the
matter density $\rho$ is decreasing function of the scale factor $a$. Minimal value of $\rho$ corresponding to the cessation
 of the expansion 
$\left(
{da}/{d\tau}= 0,\;\; a\gg 1
\right)$, and accjrding to (\ref{eq:18}),
is equal to
\begin{equation}
\rho_{\mbox{{\small min}}}=\frac{1}{16\pi G}
\left (
\frac{m_gc^2}{\hbar}
\right )^2.
\label{eq:22}
\end{equation}
If one rewrites the matter state equation in the form of (\ref{eq:1}),
then, as it is known form the first law of the thermodynamics, it follows that the  $\rho$ dependence on the scale factor $a$ will be as follows
\begin{equation}
\rho=\frac{\mbox{const}}{a^{3\nu}},
\label{eq:23}
\end{equation}
where ${\nu}=4/3$ for the relativistic matter (radiation and ``light''
neutrino) and ${\nu}=1$ for baryon matter and dark cold matter.
According to (\ref{eq:19}) the dark matter and radiation should lead to
the expansion deceleration. To explain the observed acceleration it is necessary to assume the presence of the ``dark'' energy ${\cal E}_x$ in the Universe with 
$$
\left(
\rho_x+\frac{3P_x}{c^2}
\right) < 0 \;.
$$
In this case from equation of state (\ref{eq:1}) one gets
$$
\rho_x+\frac{3P_x}{c^2}= -2\rho_x \cdot
\left(
1-\frac{3}{2}\nu
\right) \;, 
$$
and to have the acceleration it is necessary that 
$$
0 \leq \nu < \frac{2}{3}.
$$
Value of $\nu = 0$ corresponds to the presence of the vacuum energy with the density ${\cal E}_{\mbox{{\small vac}}}= \rho_{\mbox{{\small vac}}}\cdot c^2$ and $P_{\mbox{{\small vac}}}= -{\cal E}_{\mbox{{\small vac}}}$. In this case 
$\rho_{\mbox{{\small vac}}}$ does not depend on the scale factor, and
for $\rho_{\mbox{{\small vac}}}> \rho_{\mbox{{\small min}}}$ (where 
$\rho_{\mbox{{\small min}}}$ is determined by Eq. (\ref{eq:22}))
the Universe expansion, according to (\ref{eq:18}), (\ref{eq:19}), is unlimited. Thus, the relativistic theory of gravity in the Minkowsky space
is inconsistent with the presence of the vacuum energy 
$\varepsilon_{\mbox{{\small vac}}}\ne 0$. This is quite natural, since
in the flat space the density of the vacuum energy can not be different from zero. From the point of view of RTG the acceleration of the Universe expansion can be explained only by existence of quintessence (\ref{eq:1}) with parameter $\nu$ essentially greater than zero,
$$
\nu > 0. 
$$
In this case the density of the dark matter energy should decrease with the increase of the scale factor according to the law in (\ref{eq:23}), and
at values of the scale factor  large enough the existence of the nonvanishing graviton mass should lead, according to Eqs. (\ref{eq:18}) and (\ref{eq:19}), to cessation
 of the Universe expansion, which than will change into its compression. This compression, in its turn, should stop when 
some minimal value of the scale factor $a_{\mbox{{\small min}}}\ne 0$
is reached. 
Indeed, due to the fact that the left-hand side of equation (\ref{eq:18})
is positive defined,  the negative term in the right-hand side, which
increases at $a\to 0$ proportionally to $m^2_ga^{-6}$, should be compensated by the increase of the density $\rho$ (taking place in
the radiation-dominant stage and  proportional to $\rho\sim 1/a^4$). This
could happen only for $a_{\min}\ne 0$. After value
$a= a_{\mbox{{\small min}}}$ is reached, the new stage of the Universe expansion should start. Thus, the structure of the term proportional to
 $m^2_g$ in Eqs. (\ref{eq:18}) as provides the cancellation of the cosmological singularity so excludes the possibility  of the infinite
 Universe expansion. 
 In other words, according to RTG, due to the nonvanishing graviton mass the Universe evolution should proceed in the oscillating regime.
The experimental data accumulated so far allow one to estimate
the possible value of the graviton mass and basing on this value
estimate the possible oscillation period. 

\section*{\S 4. $\Omega_{\mbox{tot}}$ and graviton mass estimate}

In 1970, just after the discovery of the relic radiation, R.A. Syunyaev
and Ya.B. Zeldovich has undertook the detailed quantitative analysis of the processes taking place during the period of the hydrogen recombination and separation of the relic radiation from the matter [18]. In particular, they showed that adiabatic perturbations (sonic waves) in plasma
in the epoch of recombination should lead to the angle  anisotropy of the observed relic radiation, and studying this anisotropy one can  determine experimentally the values of some important cosmological parameters (see alsoâ earlier paper by J. Silk \cite{19} and subsequent development of this idea in \cite{20,21,22}. The question of the required accuracy 
of the angle correlation measurements for relic radiation spectrum was
discussed in details in \cite{23,24}. Among all other cosmological parameters, which values are directly determined from the 
angle characteristics of the relic radiation spectrum, there is 
the $\Omega^0_{\mbox{{\small tot}}}$ value, which is
the ratio of the total density of all matter types $(\rho)$ to current value of the critical density  $\rho^0_c$, i.e.
$$
\Omega^0_{\mbox{{\small tot}}}=\rho/\rho^0_c ~~~
\left(
\rho^0_c=3H_0^2/8\pi G\;,
\right)
$$ 
where $H_0$ is the current value of the Hubble constant \cite{25}:
$$
H_0= h \cdot (9.778\;13\cdot 10^9~\mbox{years})^{-1},
~~~ h= 0.71\pm 0.07 .
$$

In paper by A.H. Jaffe et al. \cite{26} one can find the combined analysis
of the  BOOMERANG-98 \cite{27}  and Maxima-1 \cite{28} experiments, which involves the earlier data of the  COBE~DMR \cite{29} experiments,
as well as the data resulting from the observations of the $SN1a$ supernovas \cite{1,2} and large scale structures of the Universe \cite{31}.
The results of the analysis \cite{26} show that average values of $\Omega_{\mbox{{\small tot}}}$ for combination of different experiments
systematically exceed the value 
$\Omega_{\mbox{{\small tot}}}=1$ (see Table 1, \cite{26}). 

At the confidence level of 68\% the $\Omega_{\mbox{{\small tot}}}$ value,
according to \cite{26}, is equal to
\begin{equation}
\Omega^0_{\mbox{{\small tot}}}=1.11\pm 0.07,
\label{eq:27}
\end{equation}
while, according to the inflationary theory of the early Universe
\cite{4}~$\div$ \cite{8}, the $\Omega^0_{\mbox{{\small tot}}}$ value
has to be equal to unit with a high degree of accuracy. Therefore, though the results of  \cite{26} at 95\% CL
\begin{equation}
\Omega_{\mbox{{\small tot}}}=1.11^{+0.13}_{-0.12}
\label{eq:28}
\end{equation}
do not contradict to the value of $\Omega^0_{\mbox{{\small tot}}}=1$,
the fact itself of systematical excess of average values 
$\Omega^0_{\mbox{{\small tot}}}>1$ seems to be rather designing.
Indeed, dividing both sides of Eq. (\ref{eq:18}), recalculated to the current moment, by the modern value of the Hubble constant, $H^2_0$, one can obtain the relation for (¤«ï $a\gg 1$) as follows
\begin{equation}
\Omega^0_{\mbox{{\small tot}}} = 1+f^2/6,
\label{eq:29}
\end{equation}
where $f= m_gc^{2}/\hbar H_0$. It is convenient to rewrite the $f$ parameter in the form of the ratio of the graviton mass to the 
$m^{0}_{H}$ value, which could be referred to as  {\bf the Hubble mass}:
$$
\boldmath{
m^{0}_{H} = \frac{\hbar H_0}{c^2}=
3.8\cdot 10^{-66}\cdot h~\mbox{[g]},
}
$$
\begin{equation}
f=m_g/m^0_H. 
\label{eq:31}
\end{equation}
Eqs. (\ref{eq:29})-(\ref{eq:31}) immediately give the feeling 
of the possible order of magnitude of the graviton mass. From
values (\ref{eq:28}) and expression(\ref{eq:29}) it follows that the upper limit on the graviton mass is as follows
$$
m_g\leq 1.2\cdot m^0_H ~~~ (95\% \mbox{CL}).
$$
At the same time for the confidence level of  68\%  this does not exclude that, according to (\ref{eq:27})~$\div$ (\ref{eq:29}), the graviton mass is $$
m_g=(0.8^{+0.2}_{-0.3})\cdot m^0_H
$$
Recent preliminary data of the WMAP experiment \cite{31} allow one
to make more accurate estimate on the graviton mass. According to these
data
\begin{equation}
\Omega^0_{\mbox{{\small tot}}} =1.02\pm 0.02.
\label{eq:34}
\end{equation}
From this fact it follows  that at the $(2\sigma)$ level $\;\; f^{2}/6 < 0.06$, i.e. 
$$
m_g \leq 0.6\cdot m^0_H =
1.6 \cdot 10^{-66}~\mbox{[g]}~~ (\mbox{for}~~
h = 0.71). 
$$
But within the accuracy of $(1\sigma)$ the upper limit 
$\Omega^0_{\mbox{{\small tot}}}= 1.04$ from (\ref{eq:34}) coincides with the lower value in (\ref{eq:27}). This does not exclude the possibility
\begin{equation}
\frac{f^2}{6} = 0.04 ~~ \mbox{¨} ~~
m_g\approx 0.5\cdot m^0_H = 1.3 \cdot
10^{-66}~\mbox{[g]}. 
\label{eq:36}
\end{equation}
This value of the graviton mass we will use for further estimates. 

\section*{\S5. The Universe age and the bounds on the quintessence parameter $\nu$}

As the evolution of the scale multiplier $a$ from its minimal value $a_{\min}$ to the Freedman evolution regime takes a negligible time, and the duration of the radiation-dominant stage of expansion is, at least, 
4 orders of magnitude less then the current age of the Universe, the  definition of the latter can be started immediately from the 
matter-dominant stage, assuming that the density of the cold matter  (including baryons) is equal 
$\rho_m= \rho^0_m\frac{1}{x^3}$, where $\rho^0_m$ is the current density, 
and the $x$ parameter is the ratio of the scale factor $a(\tau)$
to its current value $a_0$:
$$
x=a(\tau)/a_0.
$$
Analogously, the quintessence density can be presented in the following form
$$
\rho_q = \frac{\rho^0_q}{x^{3\nu}} ,
$$
where $\rho^0_q$ is its current value.
So, Eq. (\ref{eq:18}) takes the form
\begin{equation}
\left (
\frac{1}{x}\cdot \frac{dx}{d\tau}
\right )^2
= H_0^2 \cdot
\left(
\frac{\Omega^0_m}{x^3}
+
\frac{\Omega^0_q}{x^{3\nu}}
-\frac{f^2}{6}
\right) ,
\label{eq:38}
\end{equation}
where
$$
\Omega^0_m = \frac{\rho^0_m}{\rho^{0}_c} ~~\mbox{and}~~ \Omega^0_q
= \frac{\rho^0_q}{\rho^{0}_c} .
$$
From Eq.(\ref{eq:38}) it follows that
$$
d\tau=\frac{1}{H_0}
\frac{x^{1/2}dx}{\sqrt{F(x)} }, 
$$
where 
\begin{equation}
F(x)=
\Omega^0_m+\Omega^0_q\cdot
x^{3(1-\nu)}-\frac{f^2}{6}
x^3 .
\label{eq:40}
\end{equation}
Thus, the current age of the Universe, $t_0$, is determined by the integral 
\begin{equation}
t_0=\frac{1}{H_0}
\int\limits^{1}_0 \frac{x^{1/2}dx}{\sqrt{F(x)}},
\label{eq:41}
\end{equation}
and moments of time, which correspond to the beginning  $(t_1)$ and the cessation
  $(t_2)$ of the currently observed acceleration, are as follows
$$
t_{1(2)}=\frac{1}{H_0}
\int\limits^{x_1(x_2)}_0 \frac{x^{1/2}dx}{\sqrt{F(x)}},
$$
where $x_1$ and $x_2$ are the roots of the equation 
$$
\Omega^0_m-2\Omega^0_q
\left (1-\frac{3}{2}\nu\right )
x^{3(1-\nu)}
+\frac{f^2}{3}x^3=0,
$$
which corresponds to the zero acceleration value, see (\ref{eq:19}). The maximal time of the expansion (half-period of the oscillation, $T_0/2$) is determined by the analogous integral
$$
T_0/2=\frac{1}{H_0}
\int\limits^{x_{\max}}_0\frac{x^{1/2}dx}{\sqrt{F(x)}},
$$
where $x_{\max}$ is the root of the equation
\begin{equation}
F(x_{\max})=0. 
\label{eq:45}
\end{equation}
The Universe age  $t_0 = (13.7\pm 0.2){\cdot}10^9$ years, estimated
in \cite{31}, allows one to put bounds on the allowed region for the quintessence parameter  $\nu$ for nonvanishing value of the graviton mass.
As in the  WMAP experiment \cite{31} the value, measured directly, is  
$\omega_m = \Omega^0_m\cdot h^2= 0.135^{+0.008}_{-0.009}$, and
$\Omega^0_q = \Omega^0_\Lambda
= \Omega_{\mbox{{\small tot}}}-\Omega^0_m$
(where for the chosen graviton mass  (\ref{eq:36}) 
$\Omega_{\mbox{{\small tot}}}= 1.04$,
so the Universe age in the region of ${\bar\omega}_m-\Delta \omega_{m} <
\omega_m < {\bar \omega}_m+\Delta\omega_{m}$  acquires, according to Eqs. (\ref{eq:40})-(\ref{eq:41}), an additional dependence -- the dependence on  $h$ and $\nu$. 
Here one can determine the allowed region for the $\nu$ parameter,
which corresponds to the interval \cite{31} of the current Universe age\footnote{For distinctness we have chosen given interval for the   
Universe age as
$$
13.5\cdot 10^9\leq t_0 \leq 13.9 \cdot 10^9~\mbox{years}
$$
despite the fact that its definition 
can not be considered as model independent (see N.P. Tkachenko,
Preprint IHEP 2003).}.
This allowed parameter region is shown in Fig. 1\footnote{According to the 
PDG data \cite{25} the range for the $h$ parameter should be confined as: $0.64\leq h\leq 0.78$.}. It is interesting to note that
chosen interval for the current Universe age  requires, for
$0.64 < h < 0.67$, the existence of the quintessence with
á $\nu_{\min}> 0$. In principle, the $\nu$ value can be determined, if the higher accuracy for the $\Omega^0_m$, $\Omega^0_\Lambda$ values (assuming  $\Omega^0_\Lambda = \Omega^0_q$) and value of the acceleration $q_0$ will be achieved. According to (\ref{eq:19})
\begin{equation}
q_0 =\frac{\ddot a_0}{a_0 \cdot H^2_0}
=\left (
1-\frac{3}{2}\cdot \nu
\right ) \cdot
\Omega^0_q -\frac{\Omega^0_m}{2}
- \frac{f^2}{6} .
\label{eq:46}
\end{equation}
Excluding the ${f^2}/{6}$ value from this relation, according to
(\ref{eq:29}), one gets
\begin{equation}
\frac{3}{2} \cdot \nu \cdot \Omega^0_q =
1-q_0-\frac{3}{2}\cdot \Omega^0_m.
\label{eq:47}
\end{equation}

The recent data do not contradict to the following condition:
$$
q_0 < 1 -\frac{3}{2}\cdot \Omega^0_m ,
$$
which is necessary to fulfill the condition $\nu>0$  in Eq. (\ref{eq:47}). If one adopts for the acceleration $q_0$ the value of
$q_0=0.32\pm 0.16$, then for average values  $\Omega^0_m=0.27$ and 
$\Omega^0_q=\Omega^0_\Lambda=0.73$~[31] from Eq. (\ref{eq:47}) one gets
$\bar\nu =0.25$, and within the range of  $(1\sigma)$:  $0.05<\nu <0.43$.

\section*{\S6. Time moments corresponding to the beginning and the cessation  of the accelerated expansion epoch. Oscillation period}

Using the estimate of the possible graviton mass(\ref{eq:36}) and 
measured values of $\omega_m =\Omega^0_m \cdot h^2 = 0.135^{+0.08}_{-0.09}$ \cite{31}, one can evaluate the current Universe age and time moments 
of the beginning of the accelerated expansion $(t_1)$ and its 
cessation $(t_2)$ for different values of  $\nu$ as a function of the $h$
parameter (see Fig. 2~$\div$ 3).
In these Figs. one can see that the time of the 
accelerated expansion beginning $(t_1)$ is not too sensitive to the graviton mass  and $\nu$ parameter values, and it varies in the range of $(7\div 8)\cdot 10^9$ years. Here the smallest value of  $t_1\approx 7\cdot 10^9$ years corresponds to the largest value of the $h$ parameter, which is
compatible with the chosen interval of the Universe age. The appearance of the acceleration starting from 
$t_1\approx 7\cdot 10^9$ years  explain well-known observable paradox, 
which lies in the fact that the Hubble expansion law becomes valid
already for relatively small distances, distances of the order of few
tens Mps. (see, for instance, \cite{32}). 
With the $\nu$ increase the region of the $h$ variation, which corresponds to the chosen interval of the Universe age, shifts to region of smaller values of $h$. For example, for $\nu=0.05$ it is $0.65\leq h\leq 0.71$, but for $\nu=0.20$: $0.64\leq h\leq 0.69$.  

Time corresponding to the cessation of the accelerated expansion and the beginning of the deceleration, which leads to the cessation of the expansion, depends severely on the  $\nu$ parameter value (see Table 2). 

As it was pointed out above, the field theory of gravity in the Minlowsky space (RTG) does not allow the unlimited expansion of the Universe. Therefore,
from the point of view of RTG the only possibility
to explain the acceleration observed is the existence of the
quintessence or any other substance, which density decreases with the scale factor increase,  but not faster than $\mbox{const}/a^2$). The cessation of the expansion follows from the existence of the nonvanishing graviton mass. Under this condition the minimal value of the matter density (\ref{eq:22})
is achieved. 

The scale factor corresponding to the cessation
 of the expansion, $x_{\max}$,
is determined by the root of Eq. (\ref{eq:45}), and at small $\nu$ with a high accuracy it is equal to
$$
x_{\max}\simeq
\left(
\frac{\Omega^0_q}{{f^2}/{6}}
\right)^{1/3\nu} .
$$ 
In the given approximation it is related
with the scale factor $(x_2)$,
which corresponds to the cessation
 of the accelerated 
expansion, as follows
$$
x_2=\left(
1-\frac{3}{2}\nu
\right )^{1/3\nu}\cdot
x_{\max}\approx \frac{x_{\max}}{\sqrt{e}} .
$$
The characteristic time, corresponding to the cessation
 of the  expansion (oscillation half-period) for the chosen graviton mass (\ref{eq:36}) and for $\nu=0.05$ ia about $1300\cdot 10^9$ years, for $\nu=0.10$ it is about
$650\cdot 10^9$ years, and for $\nu=0.25$ -- about $270\cdot 10^9$ years
(see Fig. 4 and Table 2). 
It is interesting to note that the  minimal density value ($\rho_{min}$)
thus achieved  does not depend on the time of the maximal expansion ($t_{max}$)
and turns out not to be too small. Indeed, according to Eqs. (14), (18),
and (19)
\begin{displaymath}
\frac{\rho_{min}}{\rho^0_c} \simeq \frac{f^2}{6} = \Omega^0_{tot}-1\;,
\end{displaymath}
and for the chosen value of  (21)
\begin{displaymath}
\rho_{min} \simeq 0.04 \rho_c^0\;.
\end{displaymath}

The idea of the oscillatory character of the Universe evolution was
repeatedly adduced earlier, and it was stimulated mainly by the 
philosophic arguments (see, for instance, \cite{33,34}). One could expect such a regime, in principle, in the closed Freedman model with
$\Omega_{\mbox{{\small tot}}}>1$.
However, there some problem: firts, the insurmountable difficulty related with the transition through the cosmological singularity, second,
the arguments, connected with the increase of the entropy
from cycle to cycle \cite{35}.
In the RTG for the unlimited Universe the difficulties mentioned above 
are eliminated. For all that the oscillating behavior of the evolution
for infinite number of preceding cycles provides the currently observed  
average homogeneity of the matter in the Universe at large scales. 

The attraction of the Universe oscillatory evolution was stressed out in  recent paper  \cite{36}. Oscillatory regime is realized there by introducing the scalar field $\varphi$, which interacts with the matter, and using the idea of extra dimension. The authors suggest important arguments in favor of the fact that the phase of the accelerated expansion 
contributes to entropy conservation in the periodic evolution cycles. In RTG the oscillatory character of the Universe evolution is achieved
as a result of fact that the gravitational field is considered as
the physical field in the Minkowsky space and it is generated by the total energy-momentum tensor of all matter (see (\ref{eq:8}),
(\ref{eq:9})). 

In conclusion the authors would like to thank  V.V. Ezhela, V.V. Kiselev,
V.A. Petrov, P.K. Silaev,   N.E. Tyurin, and  Yu.V. Chugreev for discussion of this 
paper and valuable remarks.

\newpage
\centerline{Table 1  (from paper by A.H. Jaffe et al., [26]}
\begin{center}
\begin{tabular}{|cc|}
\hline
& $\Omega^0_{\mbox{{\small tot}}}$\\ [0.2cm] \hline
B98 +DMR & $1.15^{+0.10}_{-0.09}$\\
MAXIMA-1+DMR & $1.01^{+0.09}_{-0.09}$\\
B98+MAXIMA1+DMR & $1.11^{+0.07}_{-0.07}$\\ \hline
CMB + LSS & $1.11^{+0.05}_{-0.05}$\\ 
CMB + SN1a & $1.09^{+0.06}_{-0.05}$\\ 
CMB + SN1a+LSS & $1.06^{+0.04}_{-0.04}$\\ \hline
\end{tabular}
\end{center}

\vspace*{2cm}
\centerline{Table 2}

The time of the beginning of the Universe accelerated expansion, $t_1$,
and the time of its cessation, $t_2$. 
The time of the maximal expansion (oscillation half-period),
$t_{\mbox{{\small max}}}$, is given in billions of years. 

\begin{center}
\begin{tabular}{|r|l|l|l|}
\hline
   & \multicolumn{1}{|c|}{$t_1$} &
\multicolumn{1}{|c|}{$t_2$} &
\multicolumn{1}{|c|}{$t_{\mbox{{\small max}}}$}     \\   \hline  
$\nu = 0.05$ & 7.0 - 8.2 & 980 - 1080 & 1220 - 1360 \\  \hline 
$\nu = 0.10$ & 7.0 - 8.2 & 440 - 485  & 620 - 685   \\ \hline 
$\nu = 0.15$ & 7.1 - 8.3 & 275 - 295  & 430 - 460   \\  \hline 
$\nu = 0.20$ & 7.1 - 8.3 & 190 - 205  & 325 - 347   \\  \hline 
$\nu = 0.25$ & 7.2 - 8.5 & 142 - 149  & 263 - 280   \\  \hline 
$\nu = 0.30$ & 7.5 - 8.7 & 109 - 113  & 227 - 235   \\  \hline 
\end{tabular}
\end{center}

\newpage
\thispagestyle{empty}
\centerline{\bf Figure Captions}

\vspace*{1cm}
{\bf Fig. 1}. The range of the $\nu$ parameter variation for 
$\Omega_{\mbox{{\small tot}}}= 1.04$;
$0.126 \leq \omega_m\leq 0.143$, $13.5 < t_0 < 13.9$~GY.

\vspace*{1cm}
{\bf Fig. 2}.   The dependences of the time of the beginning of the 
Universe accelerated expansion (a) and its cessation  (b) on the $h$ value
for $\nu=0.05$ at
$\Omega_{\mbox{{\small tot}}}= 1.04; 13.5 < t_0 < 13.9$~GY;
$\nu= 0.05$; $0.126 < \omega_m < 0.143$.

\vspace*{1cm}
{\bf Fig. 3}. The same as in Fig. 2 but for $\nu = 0.20$. 

\vspace*{1cm}
{\bf Fig. 4}. The dependence of the time of the maximal
Universe expansion on the $h$ value for different values of $\nu$ at
$\Omega_{\mbox{{\small tot}}}= 1.04; 13.5 < t_0 < 13.9$~GY; 
${\nu}= 0.05$; $0.126 < \omega_m < 0.143$.

\end{document}